\documentclass[prx,superscriptaddress,twocolumn]{revtex4}
\usepackage{optidef}

\usepackage[SquareTraceBrackets]{quantum}
\usepackage{graphicx,bm,natbib,upgreek,amsmath,mathrsfs,accents}
\usepackage{amsbsy}
\usepackage[dvipsnames]{xcolor}
\definecolor{myblue}{named}{MidnightBlue}
\definecolor{mygreen}{RGB}{0,120,0}
\usepackage[T1]{fontenc}
\usepackage{newtxtext,newtxmath}

\usepackage[scaled]{helvet}
\usepackage{tikz}
\usetikzlibrary{arrows,decorations.pathmorphing,backgrounds,positioning,fit,petri}
\usepackage{gensymb}				
\usepackage[caption=false]{subfig}
\usepackage[figuresleft]{rotating}
\usepackage{siunitx}

\usepackage{lscape}

\usepackage[colorlinks=true,citecolor=myblue,linkcolor=myblue,urlcolor=myblue]{hyperref}

\DeclareMathAlphabet{\mathscrbf}{OMS}{mdugm}{b}{n} 
\DeclareMathAlphabet\mathcalbf{OMS}{cmsy}{b}{n}		

\usepackage{amsthm}

\def\>{\rangle} 
\def\<{\langle}

\DeclareGraphicsExtensions{.pdf, .jpg, .eps, .svg}

\newcommand*\dif{\mathop{}\!\mathrm{d}}

\begin{document}

\title{Quantum receiver for phase-shift keying at the single photon level}
\author{Jasminder S. Sidhu}
\email{jsmdrsidhu@gmail.com}
\affiliation{SUPA Department of Physics, The University of Strathclyde, Glasgow, G4 0NG, UK}
\affiliation{Department of Physics and Astronomy, The University of Sheffield, Sheffield, S3 7RH, UK}

\author{Shuro Izumi}
\author{Jonas  S.  Neergaard-Nielsen}
\affiliation{Center for Macroscopic Quantum States (bigQ), Department of Physics,Technical University of Denmark, Fysikvej, 2800 Kgs.~Lyngby, Denmark}

\author{Cosmo Lupo}
\email{c.lupo@sheffield.ac.uk}
\affiliation{Department of Physics and Astronomy, The University of Sheffield, Sheffield, S3 7RH, UK}

\author{Ulrik  L.  Andersen}
\affiliation{Center for Macroscopic Quantum States (bigQ), Department of Physics,Technical University of Denmark, Fysikvej, 2800 Kgs.~Lyngby, Denmark}

\date{\today}

\begin{abstract}
Quantum enhanced receivers are endowed with resources to achieve higher sensitivities than conventional technologies. For application in optical communications, they provide improved discriminatory capabilities for multiple non-orthogonal quantum states. In this work, we propose and experimentally demonstrate a new decoding scheme for quadrature phase-shift encoded signals. Our receiver surpasses the standard quantum limit and outperforms all previously known non-adaptive detectors at low input powers. Unlike existing approaches, the receiver only exploits linear optical elements and on-off photo-detection. This circumvents the requirement for challenging feed-forward operations that limit communication transmission rates and can be readily implemented with current technology.
\end{abstract}

\maketitle

\section{Introduction}
\label{sec:introduction}

\noindent
Quantum mechanics places strict fundamental limits on our ability to discriminate non-orthogonal quantum states~\cite{Helstrom1967_IC, Prosser2017_PRL}. This is a deep-rooted property of quantum mechanics which, on one hand, fuels numerous applications in quantum information science such as quantum computing and quantum key distribution~\cite{Bennett1992_PRL,Barnett2009_AOP,Bennett2014_TCS}, and on the other, limits the performance of other protocols such as sensing, metrology~\cite{Ban1997_IJTP, Jordan2015_QS, Sidhu2020_AVS,Sidhu2019_arxiv2} and communication.
The mathematical framework around state discrimination is based on the theory of quantum detection~\cite{Helstrom1967_IC}, and it has been applied to study the discrimination of various quantum states~\cite{Eldar2001_IEEE, Chou2003_PRA,Eldar2004_IEEE}.

Of particular importance is the efficient discrimination of weak coherent states.
Coherent states are endowed with an intrinsic resilience to loss and, given their immediate availability, have become indispensable information carriers in the optical realisation of classical~\cite{7174950} and quantum information protocols~\cite{PhysRevA.68.042319,RevModPhys.81.1301}. An alphabet of coherent states with very small amplitudes (down to the single photon level) possesses large state overlaps, and thus exhibits strong quantum features. Such a small-amplitude alphabet occurs often in quantum communication protocols and in classical communication schemes that aim to enhance channel capacities~\cite{PhysRevLett.92.027902}. More specifically, the optimal discrimination of weak coherent states can be used to enhance the secure key rate in quantum key distribution, improve the success rate in entanglement distillation and increase the distance of deep-space communication~\cite{4063386,7553489}.

This work focuses on the discrimination of four weak coherent states with equal amplitude and equidistant phase separations, $\{\ket{\alpha}, \ket{i\alpha}, \ket{-\alpha}, \ket{-i\alpha}\}$, chosen with equal prior probabilities, where the amplitude $\alpha$ is real-valued and positive.  This ensemble is referred to as quadrature phase shift keying (QPSK) and is commonplace in fibre networks~\cite{Garg2005_book}. It offers efficient encoding of two bits of information in one mode of the electromagnetic field. Efficient readout of the encoded information can be accomplished by measuring conjugate quadratures via a heterodyne detection~\cite{7174950,WeedbrookRMP_2012}. However, the optimal bound on the discrimination error (that is, the minimum average error in discriminating QPSK coherent states), known as the Helstrom bound, is significantly lower than that attainable through heterodyne detection~\cite{Helstrom1976,Osaki1996_PRA}. A practical setup for discriminating the QPSK coherent states at the exact Helstrom limit is unknown. However, it is possible to surpass the heterodyne limit and approach the Helstrom bound using different decoding strategies.   
These schemes generally use a combination of linear optics, photo-detection, and globally optimised displacement operations to distinguish coherent states through conditional signal nulling. 
Specifically, the average error probability can be decreased by adaptively updating the displacement phase~\cite{Bondurant1993_OL, Izumi2012_PRA,Izumi2013_PRA, Becerra2013_NP,Becerra2015_NP,PhysRevApplied.13.054015,PRXQuantum.1.010308} as well as the amplitude~\cite{Muller2015_NJP,Ferdinand2017_NPJQI}. 
Alternative sub-optimal receivers use hybrid strategies that combine homodyne detection with adaptive displacements followed by photodetection to beat the heterodyne detection limit for all signal amplitudes~\cite{Usuga2009_inproc,Muller2012_NJP}.

While receivers based on adaptive feedback indeed exhibit superior performance, e.g. outperforming the heterodyne detection limit for all amplitudes, they are technically challenging to implement and may impose practical limitations on the optical communication.
Indeed, the bandwidth of the optical communication will be intrinsically limited by the feedback mechanism since the feedback delay significantly degrades the receiver performance \cite{PhysRevApplied.13.054015}.
Alternatively, adaptive receivers can be realized by spatially dividing a signal state into multiple modes \cite{Takeoka2005_PRA,PhysRevLett.117.200501}, but this significantly increases the complexity of the receiver. 
It is therefore essential to devise a detection system that beats the heterodyne detection limit without the use of feedback techniques~\cite{Becerra2011_PRA}. It has been shown that for large coherent state amplitudes ($\alpha \gtrsim 2$), this is possible by solely using linear optics and photo-detection without the adoption of feedback~\cite{Izumi2012_PRA,DiMario:18}. However, obtaining the same advantage for weak coherent states was, up to now, an open question.

In this paper, we introduce, characterise, and experimentally demonstrate a new decoding strategy for QPSK states, comprised of 
linear optics
and on-off photo-detection. Notably, we do not make use of adaptive measurements, feed-forward, or photon number resolution. We show that adaptive feedback is not necessary to beat the conventional heterodyne decoding limit in the fully quantum, weak coherent amplitude regime ($\alpha\lesssim 0.5$). We experimentally realise the receiver and evidence strong agreement with theoretical predictions that account for the system efficiency. This work demonstrates a fundamental advance towards sub-optimal optical receivers, and provides an immediate, practical strategy to surpass the heterodyne detection limit with currently available technology. Our strategy is compatible with photon number resolving detection that can increase the robustness of the receiver against noise and extend the performance of our scheme to higher input intensities~\cite{Izumi2013_PRA, Li2013_IEE, Becerra2015_NP}.

The theoretical framework of this paper is presented in Sections~\ref{sec:coherent_state_discrimination} and~\ref{Sec:toolbox}. In Section~\ref{sec:four_coherent_states}, we present our new receiver for QPSK decoding. We demonstrate that our receiver outperforms previous decoding strategies in the weak amplitude regime, and present an experimental demonstration of this in Section~\ref{sec:experimental_demo}. Conclusions are summarised in Section~\ref{sec:conclusions}. 


\section{Theoretical framework}
\label{sec:coherent_state_discrimination}

\noindent
Consider the problem of identifying a quantum state $\rho_x$ drawn from a known finite set $\{\rho_1, \rho_2, \ldots, \rho_n\}$ with prior probabilities $\{p_1, p_2, \ldots, p_n\}$~\cite{Barnett2009_AOP,Bae2015_JPA}.
We focus on a single-shot scenario where only a single instance of the state is available. 
When the states $\rho_x$ are mutually orthogonal, detectors placed along the orthogonal directions will be able to perform perfect state discrimination. However, perfect discrimination of non-orthogonal states is not possible from a single-shot experiment and finding an optimal optical receiver is generally a difficult task. 
We consider this problem within the framework of ambiguous state discrimination, i.e., we allow for a finite probability of error that we aim to minimise.

\begin{figure}
\centering
\includegraphics[width=0.38\textwidth]{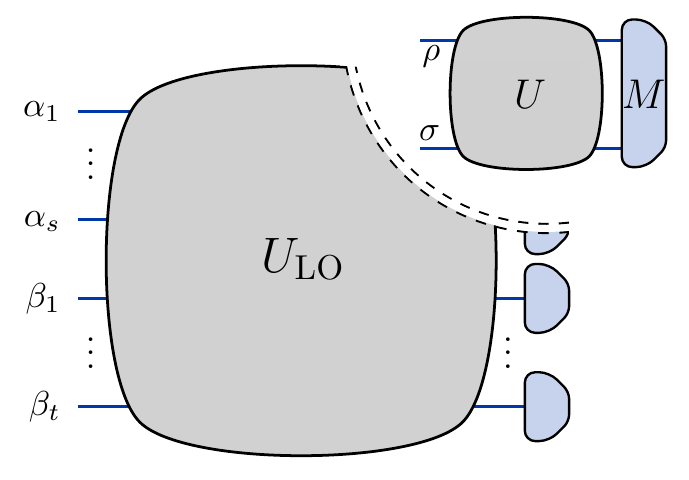}
\caption{A scheme for discrimination of multiple coherent states using linear optics and photon detection. The input ensemble is made of $s$-mode coherent states of unknown amplitudes. The scheme uses $t$ ancillary modes prepared in coherent states of known amplitudes. The final measurement is realised as $s+t$ independent photon detectors.
Inset: a general ancilla-assisted discrimination scheme, with $\rho$ an unknown state, $\sigma$ a known ancillary state, $U$ is a unitary transformation, and $M$ a measurement.} \label{fig:training}
\end{figure}

Consider a general scheme for structured detection where the unknown state $\rho$ is mixed with a known ancillary state $\sigma$ through a unitary transformation $U$. The two output systems are then measured by applying a given measurement $M$, which is characterised by the POVM elements $M_y$, with $y=\{1,\dots,m\}$. 
This is shown schematically in the inset of Fig.~\ref{fig:training}.
While the measurement is fixed, the ancillary state $\sigma$ and the unitary $U$ can be chosen within given sets, respectively denoted as $\mathcal{S}$ and $\mathcal{U}$. 

We now determine the optimal unitary $U$ and ancilla $\sigma$ that maximise the average probability of successful discrimination. 
For given $\rho_x$, $U$, and $\sigma$, the Born rule gives the probability of obtaining the measurement outcome $y$ as
\begin{align}
p_{U,\sigma}(y|x) = \mathrm{Tr} \left[ \left( U \rho_x \otimes \sigma U^\dag \right) M_y  \right] \, .
\end{align}

The Bayes rule allows us to compute the probability of input $x$ given output $y$
\begin{align}
p_{U,\sigma}(x|y) = \frac{ p_{U,\sigma}(y|x) p(x) }{p_{U,\sigma}(y)} \, .
\end{align}
The best guess for $x$, given the measurement output $y$ is the one that maximises the conditional probability,
\begin{align}
p_{U,\sigma}(\hat x |y) & = \max_x p_{U,\sigma}(x|y) \\
& = \frac{1}{p_{U,\sigma}(y)} \max_x p_{U,\sigma}(y|x) p(x)
\, ,
\end{align}
where $p_{U,\sigma}(\hat x |y)$ is the probability of successfully identifying the input state given the output measurement $y$.

The average success probability is then given by
\begin{align}
p_{U,\sigma} 
& = \sum_{y} p_{U,\sigma}(y) p_{U,\sigma}(\hat x |y) \\
& = \sum_{y} \max_x p_{U,\sigma}(y|x) p(x) \, .
\label{avepsucc}
\end{align}

%

The optimisation routine consists in finding the ancillary state $\sigma \in \mathcal{S}$ and the unitary $U \in \mathcal{U}$ that maximise $p_{U,\sigma}$. This yields the optimised success probability
\begin{align}
p_s = \sup_{U \in \mathcal{U}, \sigma \in \mathcal{S}} p_{U,\sigma} \, .
\label{eqn:opt_success_probs}
\end{align}
Note that this quantity is a function of the sets $\mathcal{U}$ and $\mathcal{S}$ only, in addition to the input ensemble and measurement $M$.
In the following section, we apply this approach to the problem of discriminating a quaternary coherent state alphabet using the linear optics toolbox. 


%
\begin{figure*}[t!]
\centering
\includegraphics[width=0.78\textwidth]{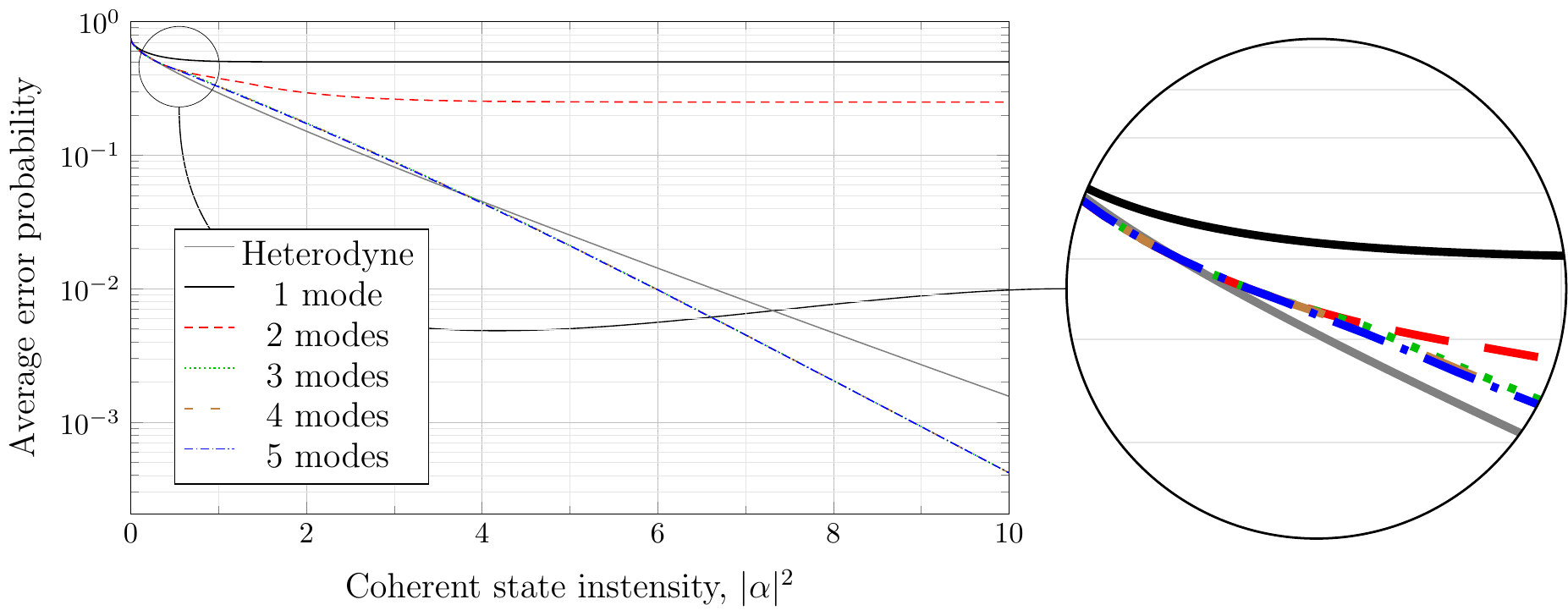}
\caption{Average error rate for QPSK signal discrimination with varying signal intensities. Here the number of modes $N$ corresponds to the dimensions of the unitary transformation of the input signal mode and $N-1$ ancillary modes. At weak photon numbers ($\abs{\alpha}^2 \lesssim 0.5$), a decoder implementing two modes (single vacuum mode) attains the smallest average error probability. For larger photon numbers ($\abs{\alpha}^2 \gtrsim 1$), three modes are optimal and sufficient to outperform Heterodyne measurements, using more than $3$ modes does not make any noticeable difference. The magnification on the RHS illustrates this clearly.} 
\label{fig:log_error_plot}
\end{figure*}%

\section{Coherent state discrimination with the linear optics toolbox}
\label{Sec:toolbox}

\noindent
Coherent state discrimination represents a concrete example of quantum state discrimination and has many applications in quantum optics. 
Here we focus on the problem of ambiguous discrimination of coherent states using linear optics and on-off photon detection.
For this, the unknown state $\rho$ is a coherent state over $s$ optical modes, $\sigma$ is a known coherent state over $t$ modes, and the measurement $M$ is mode-wise on-off photon detection. The unitary $U$ is chosen from the set of linear optical transformations over $N = s + t$ modes. While the following framework can be extended to non-classical ancillary states, we focus on coherent states given their availability and widespread use in quantum information.

It is known that linear $N$-mode unitaries can be constructed from passive linear optical transformations followed by mode-wise phase-displacement operations~\cite{Ferraro2005_arxiv,Aniello2006}. In turn, passive linear optical transformations are realised using specific arrangements of beam splitters (BS) and phase shifters~\cite{Reck1994_PRL,Clements2017}.

Consider an unknown coherent state of amplitude $\alpha_x$ ($s=1$), and $N-1$ auxiliary coherent states of amplitudes $\beta_j \in \mathbbm{C}$, for $j \in \{2,\dots, N\}$. Mixing these coherent states through an $N$-mode passive, linear optical unitary $U$, followed by mode-wise displacements $\delta_j$, yields as output on mode $j$ a coherent state with amplitude
\begin{align}
\gamma_j = U_{j1} \alpha_x + \sum_{k=2}^N U_{jk} \beta_{k-1} + \delta_j \, .
\label{eqn:general_output_amps}
\end{align}
The ancillary state amplitudes, passive linear unitary, and the displacements must be chosen to maximize the average success probability of state discrimination. The complexity of this optimisation scales quadratically with the number of modes $N$ due to the decomposition of the unitary~\cite{Reck1994_PRL,Clements2017}. We significantly reduce this complexity by noting that the amplitudes in Eq.~\eqref{eqn:general_output_amps} are also attained if the signal $\alpha_x$ is instead mixed with $N-1$ vacuum modes at the same unitary, with displacements $\smash{\epsilon_j = \sum_{k=2}^N U_{jk} \beta_{k-1} + \delta_j}$ on the $j$th mode. The original optimisation is then equivalent to requiring a general displacement $\smash{\epsilon_j \in \mathbbm{C}}$ such that $\smash{\gamma_j = U_{j1} \alpha_x + \epsilon_j}$. Notice that our use of ancillary vacuum modes renders only the first column of the unitary important, which amounts to a quadratic speedup of the optimisation process. Hence, we write 
\begin{align}
\gamma_j = u_j \alpha_x + \epsilon_j,
\label{eqn:final_state_amps}
\end{align}
where $u = (u_1, u_2, \dots, u_N)$ is a unit vector. The objective is then to determine the optimal choice of $u$ and $\epsilon$, collectively referred to as optimisation parameters, to maximise the state discrimination. 


We now define the optimised success probability for coherent state discrimination. Each mode is subject to an on-off photon detection. Denote a photon detection on the $j$th mode by $y_j=1$, and a no detection event by $y_j=0$. These mutually exclusive events occur with probabilities
\begin{align}
    p_{u_j,\epsilon_j}(y_j=1|x) & = 1 - e^{-\abs{\gamma_j}^2} 
    \, ,\\
    p_{u_j,\epsilon_j}(y_j=0|x) & =  e^{-\abs{\gamma_j}^2}
    \, ,
\end{align}
respectively. 
The overall output of the $N$-mode measurement is represented by the binary vector
$y = (y_1, y_2, \dots, y_N)$, 
%
and the average success probability in Eq.~\eqref{avepsucc} reads
\begin{align}
p_{u,\epsilon} 
& = \sum_{y} \max_x p_{u,\epsilon}(y|x) p(x) \\
& = \sum_{y} \max_x \prod_{j=1}^N p_{u_j,\epsilon_j}(y_j|x) p(x) \, .
\label{eqn:final_obj_func}
\end{align}
The latter is the objective function that we maximise by finding the optimal choice for the parameters $u$ and $\epsilon$. Note that without loss of generality, $u$ can be assumed real given the objective function is a function of the modulus of $\gamma_j$ alone. The objective function in Eq.~\eqref{eqn:final_obj_func} is then optimised over the parameters $\smash{\epsilon_j \in \mathbbm{C}}$ and $\smash{u_j \in \mathbbm{R}}$ for all $j \in \{1,2, \ldots,N\}$ with $\sum_j \abs{u_j}^2 = 1$.
This corresponds to an optimisation over a total of $3N - 1$ real parameters, which scales linearly with the number of modes.
We implement a nonlinear constrained global optimisation of the success probability. The constraints ensure that the vector $u$ has unit norm. The numerical optimiser takes advantage of primitive implementations of gradient-based and direct search algorithms for finding constrained local maximum. We have implemented these routines on Mathematica.


\section{Quadrature Phase-Shift Keying}
\label{sec:four_coherent_states}

\noindent
In QPSK, the unknown states are coherent states, $\rho _x \equiv |\alpha_x \rangle = | i^x \alpha \rangle$, with $x \in \{ 0,1,2,3 \}$, given with equal probability, and $\alpha>0$~\cite{Garg2005_book}, where $|\alpha|^2$ is the mean photon number. A practical measurement scheme to distinguish between these states is heterodyne detection, which, in average, is successful with probability
\begin{equation}\label{phet}
p_{\mathsf{het}} = \frac{1}{4} \left( 1 + \mathrm{erf}\left[\frac{\alpha}{\sqrt{2}}\right] \right)^2 \, .
\end{equation}

To apply our optimisation routine in Sec.~\ref{Sec:toolbox} to QPSK, we first fix the number of ancillary modes. 
We illustrate the achievable average error probability of distinguishing the QPSK alphabet with different number of ancillary modes in Fig.~\ref{fig:log_error_plot}. The number of modes required to minimise the error probability depends on the amplitude of the coherent states. Specifically, our numerical optimisation suggests that a decoder equipped with one ancillary coherent mode is optimal in the weak amplitude regime $\alpha \lesssim 0.5$. For larger amplitudes $\alpha \gtrsim 1$, two ancillary modes is the optimal choice. Additional ancillary modes increases the complexity of implementation while the decoding improvements over two ancillary modes are negligible. In the following, we concentrate on the weak amplitude regime and will hence consider only one ancillary mode (i.e., $N=2$).

\begin{figure}
\centering
\includegraphics[width=0.64\linewidth]{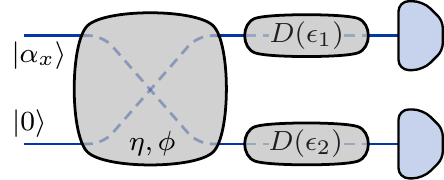}
\caption{An optimal receiver for QPSK discrimination. The unknown coherent state is first mixed with a vacuum state at a BS with transmissivity $\eta$ and phase $\phi$. Each mode is then independently displaced in phase space by $\epsilon_1$, $\epsilon_2$, before being detected using bucket detectors. This scheme is easy to implement given its independence of ancillary states and adaptive strategies.}
\label{fig:QPSK_2}
\end{figure}%

We also find an analytical solution that is an excellent approximation of the numerical optimal in the regime of weak amplitudes, with $N=2$. Furthermore, this solution does not require each parameter to be tuned to specific values of $\alpha$. This near optimal receiver is attained through: 
\begin{align}
u = \frac{1}{\sqrt{2}} \left(1, 1\right) \quad \text{and} \quad \epsilon = \frac{1}{2}\left(i+1, i-1\right) .
\end{align}
Physically, this is realised by mixing the input and ancilla modes on a $50\%$ beam splitter. The two modes are then displaced by $(i+1)/2$ and $(i-1)/2$, respectively. This is shown in Fig.~\ref{fig:QPSK_2}, where $D(\epsilon)$ denotes the phase-space displacement of amplitude $\epsilon$.
With this, the computational overheads are greatly reduced, and an experimental implementation to discriminate QPSK states below the standard quantum limit can be easily performed for weak signal amplitudes. For this near optimal choice of parameters, we obtain the following analytical expression for the average success probability (see Appendix~\ref{Sec:insight}):
\begin{align}\label{p_new}
p_\mathsf{s} = \frac{1}{4} \left( 
    1
    + 2 \, \exp\left[- \frac{1 + \alpha^2}{2}\right]
    \sinh\left[\frac{\alpha}{\sqrt{2}}\right]
    \right)^2 \, ,
\end{align}
which is close to the numerically optimised success probability. 
Note that the independence of the success probability, and hence the optimal parameters, on $\abs{\alpha}$ is only valid in the weak amplitude regime.
Intuitively, this is due to a fundamental scale in phase space that is given by the shot noise, which in our units is equal to $1$.
The optimal parameters change substantially when $\alpha \gtrsim 1$. 
In the weak amplitude regime, where 
$\alpha < 1$, these parameters remains fairly constant.

\begin{figure}[t!]
\centering
{\includegraphics[width=0.99\linewidth]{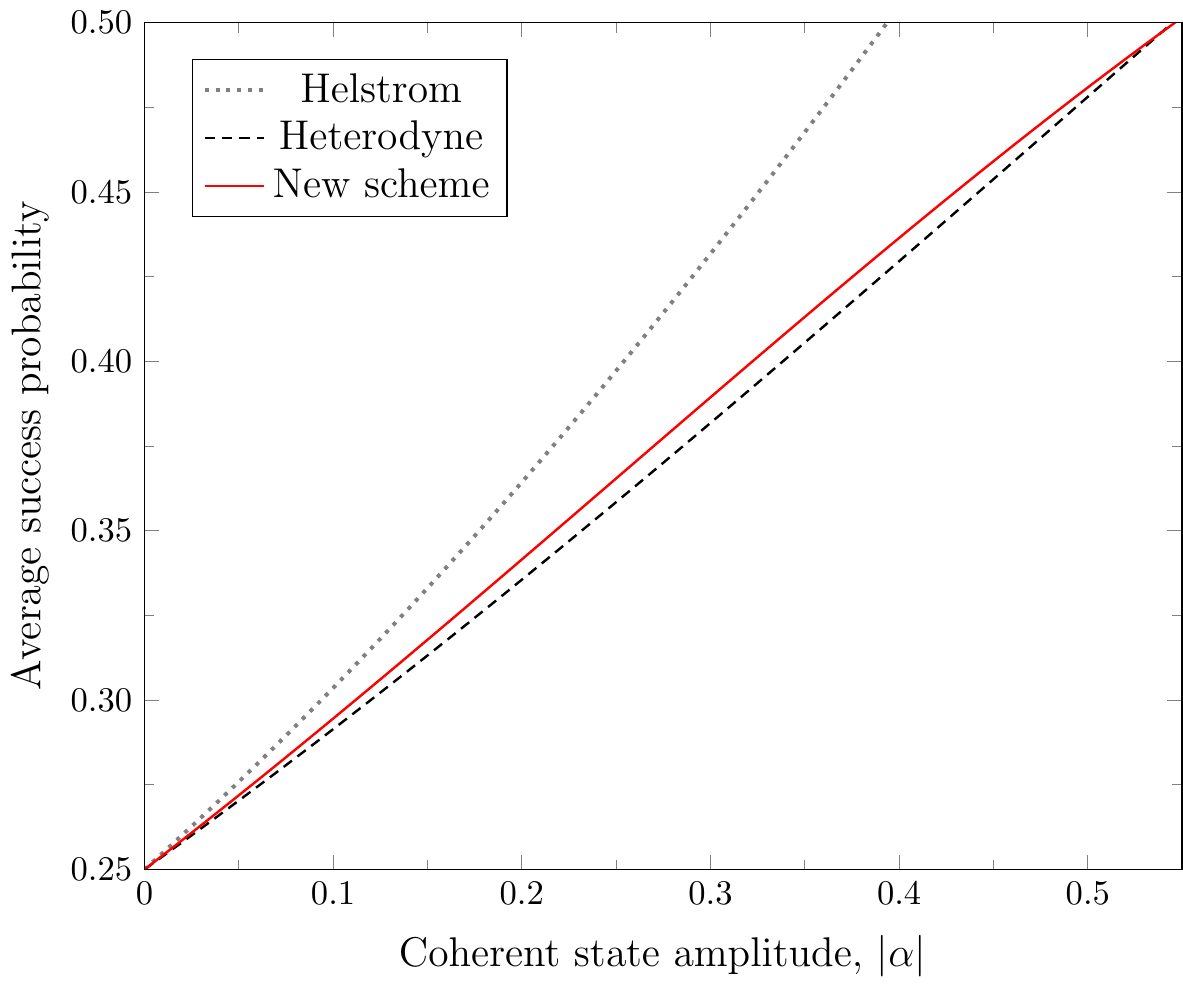}}
\caption{Optimised average success probability of distinguishing QPSK coherent states as a function of the signal amplitude $\alpha$ is shown in solid red. For comparison, the success probability attainable with heterodyne measurements is indicated with dashed black, and the Helstrom bound~\cite{Osaki1996_PRA} in dotted grey. Note that our decoder beats the SQL for weak amplitudes.}
\label{Results_1}
\end{figure}

We benchmark the success probability of our optimal receiver scheme with the Helstrom bound and heterodyne detection in Fig.~\ref{Results_1}. 
Our scheme outperforms both heterodyne detection and the hybrid scheme by M{\"u}ller \emph{et al.}~\cite{Muller2012_NJP} in the small amplitude regime. Unlike M{\"u}ller's receiver, our scheme is implemented using only linear optics and photon counting and does not rely on adaptive conditioning of the optimisation parameters.

While we have considered QPSK alphabets, our optimised routine can be applied to an arbitrary constellation of coherent states. In Appendix~\ref{sec:MPSK_recievers}, we consider the application of our detection scheme for an $M$-ary phase-shift keying code.

%

\begin{figure}[t!]
\centering
{\includegraphics[width=0.99\linewidth]{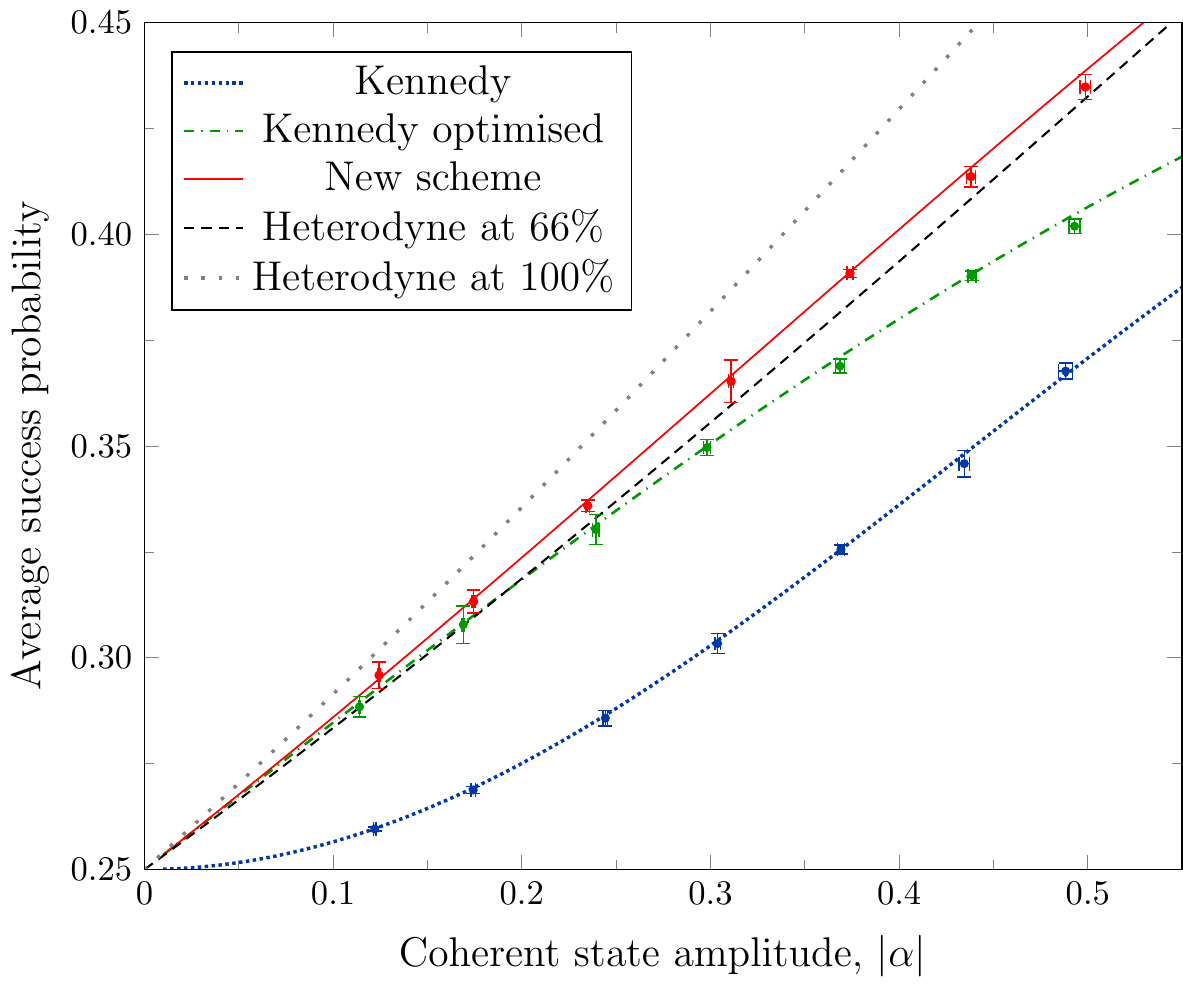}}
\caption{Average success probability of distinguishing QPSK coherent states as a function of the signal amplitude $\alpha$ for different receivers. 
The theoretical success probability of our scheme in the experimental condition is shown in solid red, together with experimental data points.
%
The success probability from heterodyne (from Eq.~(\ref{phet})) measurements accounting for $66\%$ and $100\%$ detection efficiencies are shown in dashed black and dotted grey respectively. 
The blue dotted line is for a conventional nulling Kennedy receiver~\cite{Kennedy1973_MIT}, shown together with experimental data points.
The green dash-dotted line is for the Kennedy receiver with optimised displacement amplitudes~\cite{Wittmann2008_PRL}, also shown with experimental data. Error bars denote one standard deviation from five realisations.}
\label{Results}
\end{figure}


\section{Experimental demonstration}
\label{sec:experimental_demo}

\noindent
We experimentally demonstrate our QPSK decoder in a temporal mode representation~\cite{Takeoka2005_PRA,PhysRevLett.117.200501,PhysRevLett.124.070502}.
In the spatial mode representation used thus far, beam splitters divide the signal coherent state into multiple modes while maintaining the encoded phase information. This multi-mode splitting can equivalently be accomplished in the temporal domain by splitting the signal coherent state in multiple time bins. The splitting ratio of the beam splitters corresponds to the ratio of the widths of each time bin. The displacement operations  performed individually on each mode in the spatial mode version can be implemented in the temporal mode version by instantly updating the displacement operation in time. The multiple single photon detectors in the spatial version are replaced with a {\it single} single photon detector, but it is now required that this detector has sufficiently high time resolution to detect photons in each time bin.

\begin{figure}[t!]
\centering
{
\includegraphics[width=1.00\linewidth]
{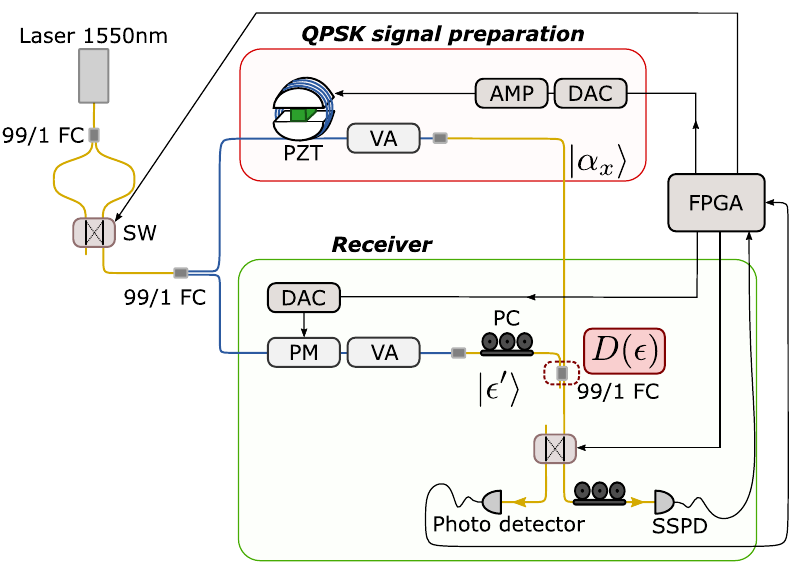}
}
\caption{Experimental setup.
FC: fiber coupler,
SW: optical switch,
PM: phase modulator,
PZT: piezo transducer,
VA: variable attenuator,
PC: polarisation controller,
SSPD: superconducting nanowire single photon detector,
DAC: digital to analog converter,
AMP: amplifier.}
\label{Experimental_setup}
\end{figure}

Our experimental setup is shown in Fig.~\ref{Experimental_setup}.
We use a continuous wave laser at \SI{1550}{\nano\metre}, which is split into two optical paths in order to individually prepare the signal coherent state and the auxiliary coherent state for the displacement operation.
A variable attenuator and a piezo transducer respectively control the amplitude and the phase of the signal state. 
A phase modulator on the auxiliary coherent state path controls the phase of the displacement operation with a maximum frequency of \SI{1}{\mega\hertz}.
Since the temporal width of the signal state is defined to be \SI{100}{\micro\second},
the displacement phase can be changed to the desired condition with little adversary effect from the finite bandwidth of the phase modulation.
The signal state is combined with the auxiliary state $\ket{\epsilon^{\prime}}$ at the 99/1 fiber coupler corresponding to the physical implementation of the displacement operation.
Using an optical switch, the interfered beam is guided to either a photo detector for the purpose of stabilising the relative phase between signal and auxiliary states, or a superconducting nanowire single photon detector (SSPD) for data acquisition.
Because the conventional photon detector cannot measure the laser power highly attenuated to photon level,
the laser power is also switched between high and low by an optical switch after the laser source.
A field programmable gate array (FPGA) collects the electrical signals from the SSPD and generates the signal driving the phase modulator.
We achieve a total system efficiency of about $66\%$, where the transmission efficiency from before the 99/1 fiber coupler to the SSPD is approximately $90\%$. The detection efficiency of the SSPD is measured to be approximately $73\%$ and the dark count noise around \SI{25}{\hertz}, which corresponds to $2.5\times10^{-3}$ counts per signal~\cite{PhysRevApplied.13.054015}. The dark count noise, as well as the non-unit visibility, are critical experimental imperfections that limit the performance of receivers based on  displacement and photon detection~\cite{Izumi2013_PRA,Becerra2015_NP}. 
Nevertheless, since we demonstrate our strategy with very weak coherent amplitude conditions, the error probability is so high that the contribution of the error induced by the dark count noise and the visibility imperfection is negligibly small. 

\begin{figure*}[ht!]
\centering
{\includegraphics[width=0.85\textwidth]{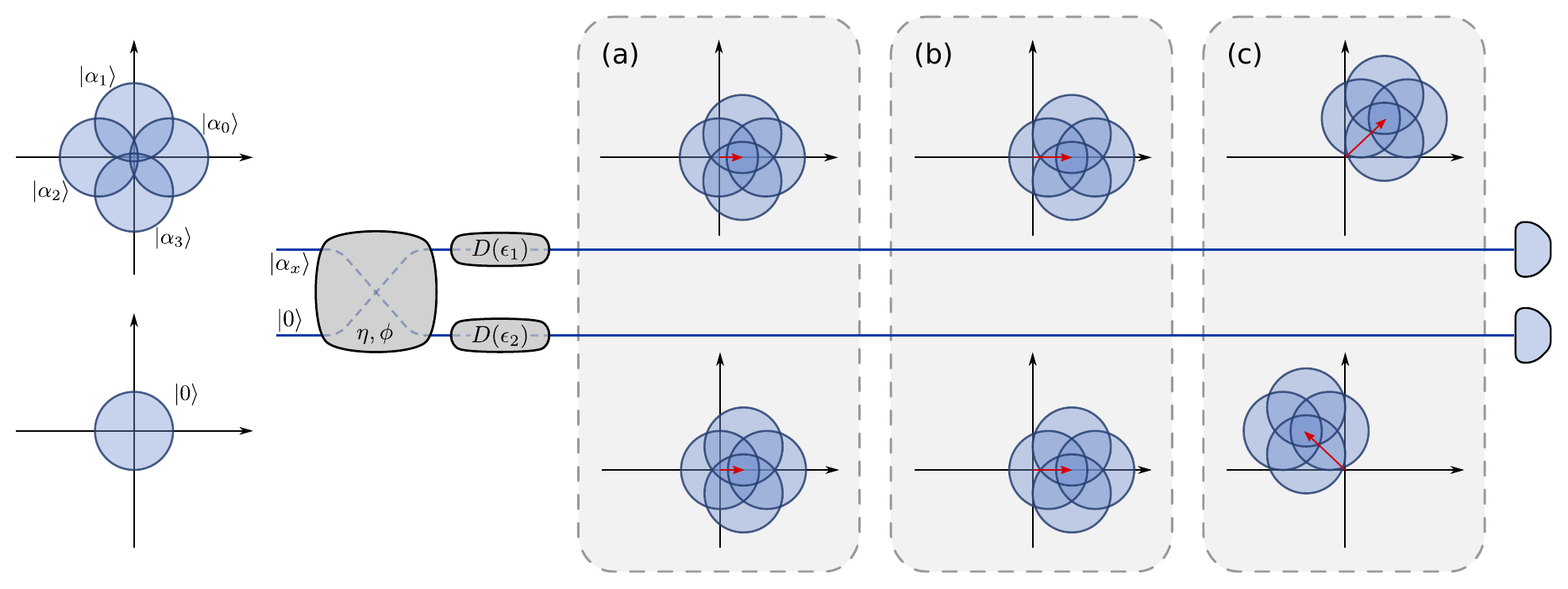}}
\caption{Illustration of the three receiver schemes implemented in the experiment for discriminating between the four input QPSK states $|\alpha_0\rangle, |\alpha_1\rangle, |\alpha_2\rangle, |\alpha_3\rangle$. 
(a) In the nulling Kennedy receiver~\cite{Kennedy1973_MIT}, both BS outputs are displaced such that the $|\alpha_2\rangle$ state is shifted to the phase space origin. (b) The amplitude-optimised Kennedy receiver~\cite{Wittmann2008_PRL} is similar to the nulling receiver, but by displacing $|\alpha_2\rangle$ further past the origin, the success probabilities for weak amplitudes are significantly improved. (c) Our optimal receiver also optimises the phases of the displacements on the two BS outputs, leading to further improvements. }
\label{fig:phasespace_displacements}
\end{figure*}

We experimentally investigate the performance of three types of two-mode receivers based on displacement operations and photon detections.
The three schemes are illustrated on phase space diagrams in Fig.~\ref{fig:phasespace_displacements} and the experimentally obtained performances of the receivers are depicted in Fig.~\ref{Results}.
(1) A conventional nulling Kennedy receiver which implements the displacement operations such that one of the QPSK signals is displaced to the vacuum state, i.e. $\smash{\abs{\epsilon_1}=\abs{\epsilon_2}=\abs{\alpha}/\sqrt{2}}$ with $\eta=1/2$~\cite{Kennedy1973_MIT} (Fig.~\ref{fig:phasespace_displacements}(a)).
The mean and error bars of the success probabilities are evaluated from five independent procedures with $4\times10^4$ data points for each procedure. The signal amplitude is calibrated from the observed photon count rate by blocking the auxiliary state path and the error bars are evaluated from ten independent procedures.
(2) The Kennedy receiver with optimised displacement amplitude~\cite{Wittmann2008_PRL} (Fig.~\ref{fig:phasespace_displacements}(b)).
Our numerical analysis indicates that, for both nulling and displacement amplitude optimised receivers, the displacement phases for modes 1 and 2 should be set to the same value to maximise the average success probability in the very weak amplitude case.
For the displacement amplitude optimised receiver, the near optimal performance for the weak coherent signal amplitude is obtained with $\abs{\epsilon_1}=\abs{\epsilon_2}=1/2$ and $\eta=1/2$.
The conventional approaches are unable to beat the heterodyne limit, given by the black dash-dotted line, in the weak coherent amplitude range.
(3) Our strategy of optimising the phase of the displacement operations  (Fig.~\ref{fig:phasespace_displacements}(c)), implemented with the near optimal parameters, provides an improved performance that overcomes the heterodyne limit.
Since our system has the finite detection efficiency of $66\%$, the success probability is degraded and the performances in the actual experimental condition are plotted for the optimal receiver by a dashed curve and open circles for theory and experiment, respectively.

\section{Conclusions and discussions}\label{sec:conclusions}

\noindent
Non-orthogonal quantum states are the building blocks of quantum communication protocols. 
Weak coherent states are commonly used in these applications given their non-orthogonality, relative ease of generation in laboratories, and resilience to loss.
Motivated by this, we have looked at designing practical receivers to discriminate coherent states. In particular, we have focused on the optimal receivers that can be obtained by combining linear optics (including passive linear optics and phase-space displacements) and on-off photo-detectors.

The natural decoders for discriminating coherent states are homodyne and heterodyne detections. These detectors have the advantage of being already commonly employed in standard telecommunications. However, they are limited by the shot noise. Several works have focused on the design of structured receiver that could beat the shot noise limit. In particular, for the problem of decoding QPSK, the only known sub-shot-noise strategies in the low amplitude regime exploit feed-forward~\cite{Muller2012_NJP,Ferdinand2017_NPJQI, Muller2015_NJP}. Although feasible in principle, approaches based on feed-forward remain technically demanding.

Furthermore, the delay due to practical factors associated with the feedback, such as the signal propagation time, the speed of signal processing
after photon detection and the response time of the single photon detector, and the delay of updating the
displacement operation may significantly degrade the error probability \cite{PhysRevApplied.13.054015}.
The delay becomes more critical if the temporal extent of the signal carrier becomes shorter and therefore the bandwidth of the feedback would limit the possible bandwidth of the optical communication. On the other hand, since static strategies without feedback control do not require real time signal processing, our receiver is compatible with the high speed optical communication where the temporal width of the signal carrier is intrinsically short.

Here, we develop a novel sub-shot-noise QPSK decoding. Our scheme employs linear optics 
and on-off photo-detectors, without feed-forward operations. We demonstrate that it outperforms heterodyne detection, as well as all previous non-adaptive detectors in the weak pulse regime. 
Experimental implementation of our 
receiver demonstrates results consistent with our theoretical analysis. 
Going beyond QPSK, we have shown that our scheme allows us to beat heterodyne detection for $3$-PSK decoding, but not for $5$-PSK and $6$-PSK (Appendix \ref{sec:MPSK_recievers}). This suggests that our scheme can beat heterodyne in $M$-PSK detection only for $M \leq 4$.

An interesting experimental platform to test some of this work is the hybrid spatio-temporal architecture for universal linear optics~\cite{Su2019_PRA}. This scheme would be useful to implement the optimised unitary receivers that we construct based on the design by Reck, Zeilinger, Bernstein, and Bertani~\cite{Reck1994_PRL}. It would be interesting to see how this proposal compares with the experimental minimum error measurements proposed by Sol{\'s}-Prosser \emph{et al.}~\cite{Prosser2017_PRL}.

Our results pave the way for a number of research questions. 
First, photon number resolving detectors may further improve our decoding strategy, especially in the region of higher signal amplitudes. Second, the effects of non-classical ancillary states are also expected to deliver improvements to the attainable success probability~\cite{Nair2012_PRA}. The extent of improvement to our scheme is an interesting line of future work.
%
Third, though we have focused on ambiguous state discrimination, our approach may also be useful 
for unambiguous discrimination of coherent states~\cite{Peres_1998}.
Finally, it may be combined with error correcting codes and exploited to demonstrate the phenomenon of super-additivity in quantum communication \cite{Guha2011_PRL}.

\section*{Acknowledgements}

\noindent
JSS and CL acknowledge EPSRC for funding via the Quantum Communications Hub (EP/M013472/1).
SI, JSNN and ULA acknowledge the Danish National Research Foundation through the Center for Macroscopic Quantum States (DNRF142).
We would like to thank Masahiro Takeoka and Saikat Guha for useful discussions on an earlier version of the manuscript and T. Yamashita, S. Miki, and H. Terai for providing and installing the superconducting nanowire single photon detector.

\section*{Author contributions}

\noindent
JSS performed the numerical optimisation and wrote the initial version of the manuscript. CL designed the project and steered the direction of research. SI, JSN, and ULA designed the experiment and wrote about this, while SI performed the experiment and analysed the data. All authors contributed equally in selecting relevant literature and checking the manuscript for accuracy. 

\bibliographystyle{ieeetr}



\appendix

\section{Almost optimal discrimination of QPSK}\label{Sec:insight}

\noindent
Consider the receiver in Fig.~\ref{fig:QPSK_2}, where the BS has $50\%$ transmissivity and zero phase, and the complex displacements on both modes have amplitudes $\epsilon_1 = (i+1)/2$,
$\epsilon_2 = (i-1)/2$.

For a given $x$, the action of the BS maps the input and ancillary coherent state in two coherent states of amplitudes:
\begin{align}
\begin{split}
\gamma_+ & = \frac{1}{\sqrt{2}} \, e^{i x \pi/2 } \alpha  + \frac{i+1}{2} \, , \\
\gamma_- & = \frac{1}{\sqrt{2}} \, e^{i x \pi/2 } \alpha + \frac{i-1}{2} \, .
\end{split}
\end{align}

Defining $p(y|x)$ as the probability of obtaining the measurement outcome $y$ given the input state $\alpha_x$, then the probability of individual detectors to click is given by
\begin{align}
p_\pm(1|x)
& = 1 - e^{-|\gamma_\pm|^2} \\
& = 1 - \exp 
\left[
- \frac{1+\alpha^2}{2} - \frac{\alpha}{\sqrt{2}} \left\{ \sin\left(\frac{x \pi}{2}\right) \pm \cos\left(\frac{x \pi}{2}\right) \right\}
\right] \, .
\end{align}

In conclusion, we table the detection probability for each input state:
\begin{align}
x=0 & \to 
\left\{
\begin{array}{l}  
p_+(1|0) = 1 - \exp 
\left[
- \frac{1 + \alpha^2 + \sqrt{2}\alpha}{2} 
\right] \vspace{2pt}\\
p_-(1|0) = 1 - \exp 
\left[
- \frac{1 + \alpha^2 - \sqrt{2}\alpha}{2}
\right]
\end{array} 
\right. \\
x=1 & \to 
\left\{
\begin{array}{l}  
p_+(1|1) = 1 - \exp 
\left[
- \frac{1 + \alpha^2 + \sqrt{2}\alpha}{2}
\right] \vspace{2pt}\\
p_-(1|1) = 1 - \exp 
\left[
- \frac{1 + \alpha^2 + \sqrt{2}\alpha}{2}
\right]
\end{array}
\right. \\
x=2 & \to 
\left\{
\begin{array}{l}  
p_+(1|2) = 1 - \exp 
\left[
- \frac{1 + \alpha^2 - \sqrt{2}\alpha}{2} 
\right] \vspace{2pt}\\
p_-(1|2) = 1 - \exp 
\left[
- \frac{1 + \alpha^2 + \sqrt{2}\alpha}{2}
\right]
\end{array}
\right. \\
x=3 & \to 
\left\{
\begin{array}{l}  
p_+(1|3) = 1 - \exp 
\left[
- \frac{1 + \alpha^2 - \sqrt{2}\alpha}{2}
\right] \vspace{2pt}\\
p_-(1|3) = 1 - \exp 
\left[
- \frac{1 + \alpha^2 - \sqrt{2}\alpha}{2}
\right]
\end{array}
\right. 
\end{align}
From this table, we compute the maximum likelihood estimation for each combination of detection events, and the associated average probability of successful state discrimination:
\begin{align}
    p_s = \frac{1}{4} \left( 
    1
    + 2 \, \exp\left[- \frac{1 + \alpha^2}{2} \right] 
    \sinh{ \frac{\alpha}{\sqrt{2}}}
    \right)^2 \, .
\label{eqn:approx_analytic_sol}    
\end{align}
Hence, we have found a fixed receiver setup, where the displacements are independent of $\abs{\alpha}$.
Note that the independence of the success probability, and hence the optimal parameters, on $\abs{\alpha}$ is only valid in the weak amplitude regime.

\begin{figure}[t!]
\centering
\includegraphics[width=\columnwidth]{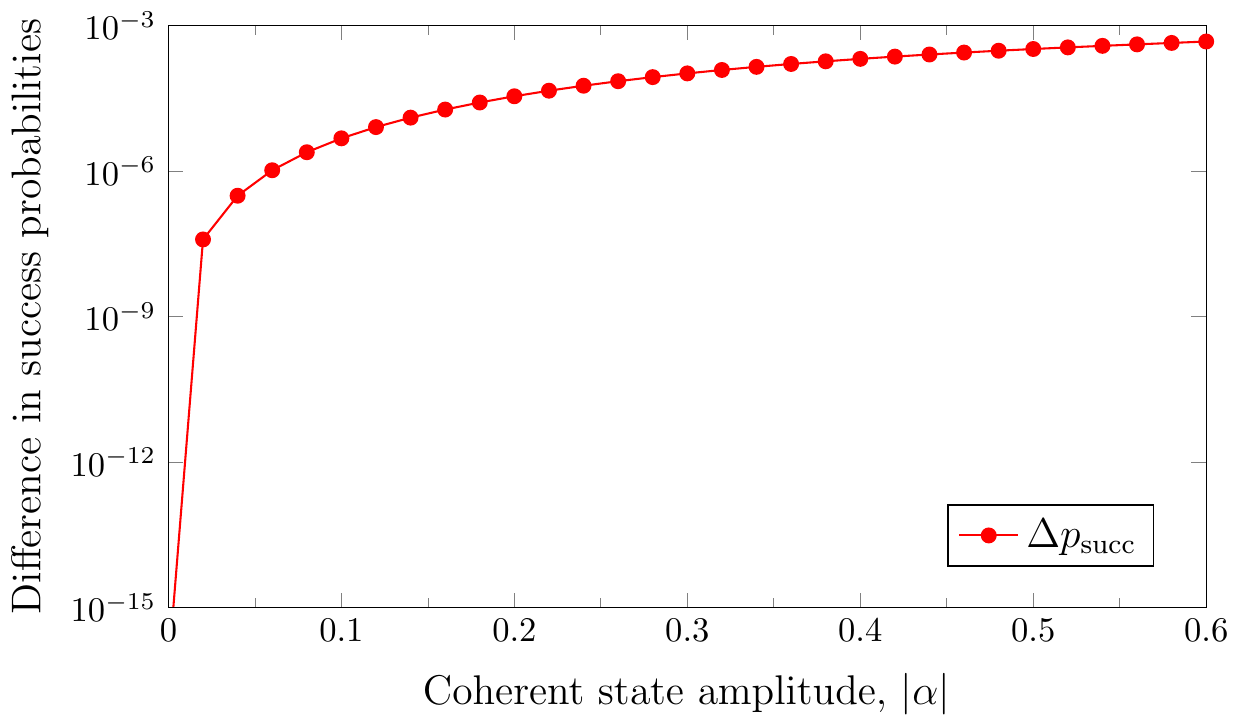}
\caption{Difference between the numerically optimised success probability of discriminating QPSK constellations, with the corresponding approximate solution in Eq.~\eqref{eqn:approx_analytic_sol}. We note that in the weak signal regime, our analytic solution is a good approximate to the optimised solution.}
\label{fig:diff_succ_sols}
\end{figure}

Figure~\ref{fig:diff_succ_sols} illustrates the difference between the numerically optimised success probability and the approximate analytic solution in Eq.~\eqref{eqn:approx_analytic_sol}. Notice that the difference is negligible and increases as the signal amplitudes approach the classical regime ($\alpha \gtrsim 1$), where the receiver is no longer independent of $\alpha$. Specifically, in this regime, vacuum fluctuations have minimal effect and any change to signal amplitudes effect large changes in the attainable success probability. This highlights that it is possible to implement a signal amplitude independent receiver and still obtain a good approximate to the fully optimised receiver, provided $\abs{\alpha} \lesssim 1$.

%


\section{M-PSK receivers}
\label{sec:MPSK_recievers}

\noindent
In this section, we explore the application of our detection scheme for an $M$-PSK ($M$-ary phase-shift keying) code.
For arbitrary $M$ this code is defined by $M$ equally likely coherent states with amplitudes $\alpha_k = \alpha e^{i k 2\pi /M}$, for $k=\{1,2,\ldots, M\}$.
We compare our scheme with heterodyne detection, which for $M$-PSK yields the average success probability (see Appendix~\ref{Sec:het})
\begin{align}
\begin{split}
p_\mathrm{het} = &\frac{1}{M} \, \exp\left[ - \alpha^2 \right]
+ \frac{1}{2} \,
\mathrm{erf}\left[ \alpha \sin\left[\frac{\pi}{M}\right]\right] \\
& + \frac{1}{\sqrt{\pi}} 
\int_{0}^{ \alpha \sin\left[\frac{\pi}{M}\right]} \dif u \;
\exp\left[ - u^2 \right]
\mathrm{erf}\left[  \sqrt{ \alpha^2 - u^2} \right] \, .
\end{split}
\end{align}

\begin{figure}[t!]
\centering
\includegraphics[width=0.5\linewidth]{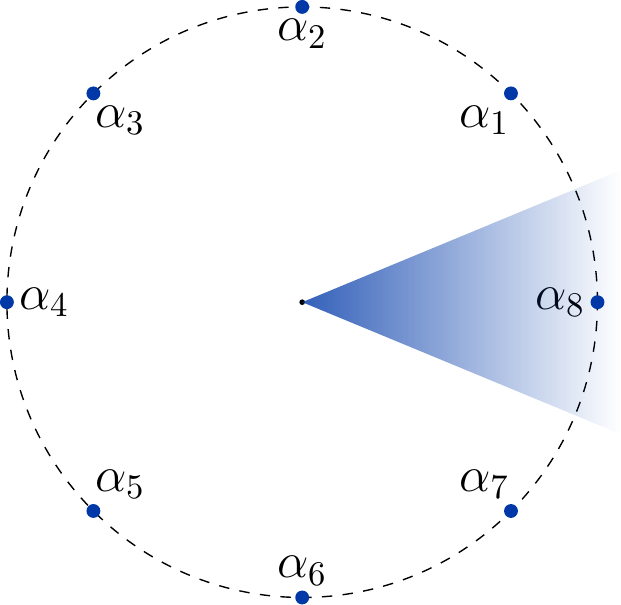}
\caption{A constellation of $M$ coherent states distributed with phase separation $2\pi/M$ with $M=8$.}
\label{fig:constellation}
\end{figure}%

We have considered the case of $M = 3,5,6$, and optimised our detection scheme using variable number of modes $N$.
Fig.~\ref{fig:MPSK_Nmodes} illustrates the effect of increasing the number of ancillary modes on the average error probability of the optimal schemes. 
The average error probability of discriminating an $M$-PSK alphabet using heterodyne detection is also illustrated for comparison. The discrimination capability of heterodyne detection increases with increasing $M$.

In general, there is an advantage in using more ancillary modes. However, the average error probability quickly saturates, especially when the number of coherent states, $M$, is small.
Specifically, for $M=3$ there is no noticeable improvement in using more than one ancilla for all signal amplitudes. Similarly, for $M=4$, there is no improvement in using more than two ancillas for all signal amplitudes (see Fig.~\ref{fig:log_error_plot}).
%
The use of more ancillas appears to be more beneficial as $M$ increases. However, our scheme does not beat heterodyne detection for $M = 5, 6$. This suggests that the standard quantum limit can be exceeded only for $M = 3,4$ when using linear passive optics, without feedback.


\section{M-PSK average success probability of heterodyne detection} \label{Sec:het}

\noindent
In this section, we derive an expression for the success probability of distinguishing coherent states from the $M$-PSK alphabet. This is a set of $M$ coherent states with amplitudes $\alpha_k = \alpha e^{i 2 \pi k /M}$. The $M$-PSK coherent states are represented in phase space in Fig.~\ref{fig:constellation}. 
Without loss of generality, we assume $\alpha$ real and put $\alpha = x_0 /\sqrt{2}$.
Measuring by heterodyne detection a single coherent state with amplitude $\alpha = \frac{x_0}{\sqrt{2}}$ leads an outcome $(x,y)$ with probability density distribution:
\begin{align}
P(x,y) = \mathcal{N} \exp{ \left[ - \frac{(x-x_0)^2 + y^2 }{2\sigma^2} \right]} \, ,
\end{align}
where $\mathcal{N} = (2\pi \sigma^2)^{-1}$, and $\sigma^2 = 1$ is the shot-noise (variance of the vacuum).
We need to express this in polar coordinates:
\begin{align}
P(r,\theta) = \mathcal{N} \exp{ \left[ -\frac{r^2 + x_0^2 - 2 r x_0 \cos{\theta} }{2\sigma^2} \right]} \, ,
\end{align}
where
\begin{align}
r = \sqrt{ x^2 + y^2} \, , \, \,
\cos{\theta} = \frac{x}{r} \, .
\end{align}

Consider an ensemble of $M$ coherent states with amplitudes $\alpha_k = \frac{x_0}{\sqrt{2}} \, e^{i 2 k \pi/M}$.
If these states have equally likely, then the probability of correct guessing from the output of heterodyne detection is
\begin{align}
p_\mathrm{het} & =  \int_{-\pi/M}^{\pi/M} d\theta
\int_0^\infty r \dif r \;
P(r,\theta) \\
& = \mathcal{N} e^{ - \frac{x_0^2}{2\sigma^2} }
\int_{-\pi/M}^{\pi/M} \dif\theta \;
\int_0^\infty r \dif r \;
 \exp{ \left[ -\frac{r^2 - 2 r x_0 \cos{\theta} }{2\sigma^2} \right]}  \, .
\end{align}
This corresponds to the integral over the colored region in Fig.\ \ref{fig:constellation}. We can simplify the integral with the change of variable
\begin{align}
\rho := \frac{r^2 - 2r x_0 \cos{\theta}}{2\sigma^2} \, .
\end{align}

We have
\begin{widetext}
\begin{align}
\begin{split}
p_\mathrm{het} 
%
& = \sigma^2 \mathcal{N} e^{ - \frac{x_0^2}{2\sigma^2} }
\int_{-\pi/M}^{\pi/M} \dif \theta \;
\int_0^\infty \dif\rho \; e^{-\rho}
+ x_0 \mathcal{N} e^{ - \frac{x_0^2}{2\sigma^2} }
\int_{-\pi/M}^{\pi/M} \dif\theta \; \cos{\theta}
\int_0^\infty \dif r \; 
 \exp{ \left[ -\frac{r^2 - 2 r x_0 \cos{\theta} }{2\sigma^2} \right]} \\
& = \sigma^2 \mathcal{N} e^{ - \frac{x_0^2}{2\sigma^2} }
\frac{2\pi}{M}
+ x_0 \mathcal{N} 
\int_{-\pi/M}^{\pi/M} \dif\theta \; \cos{\theta}
\exp\left[ - \frac{x_0^2}{2\sigma^2}\left(1-(\cos{\theta})^2\right) \right]
\int_0^\infty \dif r \; 
 \exp{ \left[ -\frac{ ( r - x_0 \cos{\theta})^2 }{2\sigma^2} \right]} \, .
\end{split}
\end{align}

Now we substitute $\smash{\mathcal{N} = (2\pi \sigma^2)^{-1}}$ and make the following change of variables:
\begin{align}
Y := \sin{\theta} \, , \, \, 
r' := \frac{r - x_0 \cos{\theta}}{\sqrt{2} \sigma} \, .
\end{align}

This yields
\begin{align}
\begin{split}
p_\mathrm{het} 
& = \frac{1}{M} \, \exp\left[ - \frac{x_0^2}{2\sigma^2} \right]
+  \frac{x_0}{2\pi \sigma^2}
\int_{-\pi/M}^{\pi/M} \dif\theta \; \cos{\theta}
\exp\left[ - \frac{x_0^2}{2\sigma^2}[1-(\cos{\theta})^2] \right]
\int_0^\infty \dif r \; 
 \exp{ \left[ -\frac{ ( r - x_0 \cos{\theta})^2 }{2\sigma^2} \right]} \\
& = \frac{1}{M} \, \exp\left[ - \frac{x_0^2}{2\sigma^2} \right]
+   \sqrt{ \frac{\pi}{2} } \frac{x_0}{2\pi \sigma}
\int_{-\sin{\pi/M}}^{\sin{\pi/M}} \dif Y \;
\exp\left[ - \frac{x_0^2}{2\sigma^2} Y^2 \right]
 \frac{2}{\sqrt{\pi}}
\int_{-\frac{x_0 \cos{\theta}}{\sqrt{2}\sigma}}^\infty \dif r' \; e^{-{r'}^2} \\
& = \frac{1}{M} \, \exp\left[ - \frac{x_0^2}{2\sigma^2} \right]
+  \sqrt{\frac{\pi}{2}} \frac{x_0}{2\pi \sigma}
\int_{-\sin{\pi/M}}^{\sin{\pi/M}} \dif Y \;
\exp\left[ - \frac{x_0^2}{2\sigma^2} Y^2 \right]
\left( 1 + \mathrm{erf}\left[ \frac{x_0 \sqrt{1-Y^2}}{\sqrt{2}\sigma}\right] \right) \, ,
\end{split}
\end{align}
where in the last equation we have applied the identities
\begin{align}
\frac{2}{\sqrt{\pi}} \int_{u}^\infty \dif r' \; \exp\left[-{r'}^2\right] = 1 -  \frac{2}{\sqrt{\pi}} \int_{0}^u \dif r' \; \exp\left[-{r'}^2\right]
= 1 - \mathrm{erf}[u] 
= 1 + \mathrm{erf}[-u] \, .
\end{align}

We obtain
\begin{align}
p_\mathrm{het} 
& = \frac{1}{M} \, \exp\left[ - \frac{x_0^2}{2\sigma^2} \right]
+  \sqrt{\frac{\pi}{2}} \frac{x_0}{2\pi \sigma}
\int_{-\sin{\pi/M}}^{\sin{\pi/M}} \dif Y \;
\exp\left[ - \frac{x_0^2}{2\sigma^2} Y^2 \right]
+  \sqrt{\frac{\pi}{2}} \frac{x_0}{2\pi \sigma}
\int_{-\sin{\pi/M}}^{\sin{\pi/M}} \dif Y \;
\exp\left[ - \frac{x_0^2}{2\sigma^2} Y^2 \right]
\mathrm{erf}\left[ \frac{x_0 \sqrt{1-Y^2}}{\sqrt{2}\sigma}\right]\nonumber
\\
& = \frac{1}{M} \, e^{ - \frac{x_0^2}{2\sigma^2} }
+ \sqrt{\frac{\pi}{2}} \frac{x_0}{\pi \sigma}
\int_{0}^{\sin{\pi/M}} \dif Y \;
e^{ - \frac{x_0^2}{2\sigma^2} Y^2 }
+ \sqrt{\frac{\pi}{2}} \frac{x_0}{\pi \sigma}
\int_{0}^{\sin{\pi/M}} \dif Y \;
\exp\left[ - \frac{x_0^2}{2\sigma^2} Y^2 \right]
\mathrm{erf}\left[ \frac{x_0 \sqrt{1-Y^2}}{\sqrt{2}\sigma}\right] \, .
\end{align}

By applying the following change of variables
\begin{align}
Y' := \frac{x_0}{\sqrt{2}\sigma} Y
\end{align}
we obtain
\begin{align}
\begin{split}
p_\mathrm{het} 
& = \frac{1}{M} \, \exp\left[ - \frac{x_0^2}{2\sigma^2} \right]
+ \frac{1}{2} \frac{2}{\sqrt{\pi}}
\int_{0}^{\frac{x_0}{\sqrt{2}\sigma} \sin{\pi/M}} \dif Y' \;
e^{ - {Y'}^2 }
+ \frac{1}{\sqrt{2\pi}} \frac{x_0}{\sigma}
\int_{0}^{\sin{\pi/M}} \dif Y \; 
\exp\left[ - \frac{x_0^2}{2\sigma^2} Y^2 \right]
\mathrm{erf}\left[ \frac{x_0 \sqrt{1-Y^2}}{\sqrt{2}\sigma}\right] \\
& = \frac{1}{M} \, \exp\left[ - \frac{x_0^2}{2\sigma^2} \right]
+ \frac{1}{2} \,
\mathrm{erf}\left[ \frac{x_0}{\sqrt{2}\sigma}\sin{\frac{\pi}{M}}\right]
+ \frac{1}{\sqrt{\pi}} 
\int_{0}^{\frac{x_0}{\sqrt{2}\sigma}\sin{\pi/M}} \dif Y'
\; \exp\left[ - {Y'}^2 \right]
\mathrm{erf}\left[  \sqrt{\frac{x_0^2}{2\sigma^2}-{Y'}^2} \right] \, .
\label{integ0}
\end{split}
\end{align}

Finally, putting $x_0 = \alpha \sqrt{2}$ and $\sigma^2=1$, we obtain
\begin{align}
p_\mathrm{het} 
= \frac{1}{M} \, e^{ - \alpha^2 }
+ \frac{1}{2} \,
\mathrm{erf}\left[ \alpha \sin{\frac{\pi}{M}}\right]
+ \frac{1}{\sqrt{\pi}} 
\int_{0}^{ \alpha \sin{\pi/M}} \dif Y' \;
\exp\left[ - {Y'}^2 \right]
\mathrm{erf}\left[  \sqrt{ \alpha^2 - {Y'}^2} \right] \, ,
\end{align}
which concludes our proof for the average success probability of distinguishing an $M$-PSK alphabet using heterodyne measurements.


%
\begin{figure*}[t!]
\subfloat[3-PSK.]{\includegraphics[width=0.326\linewidth]{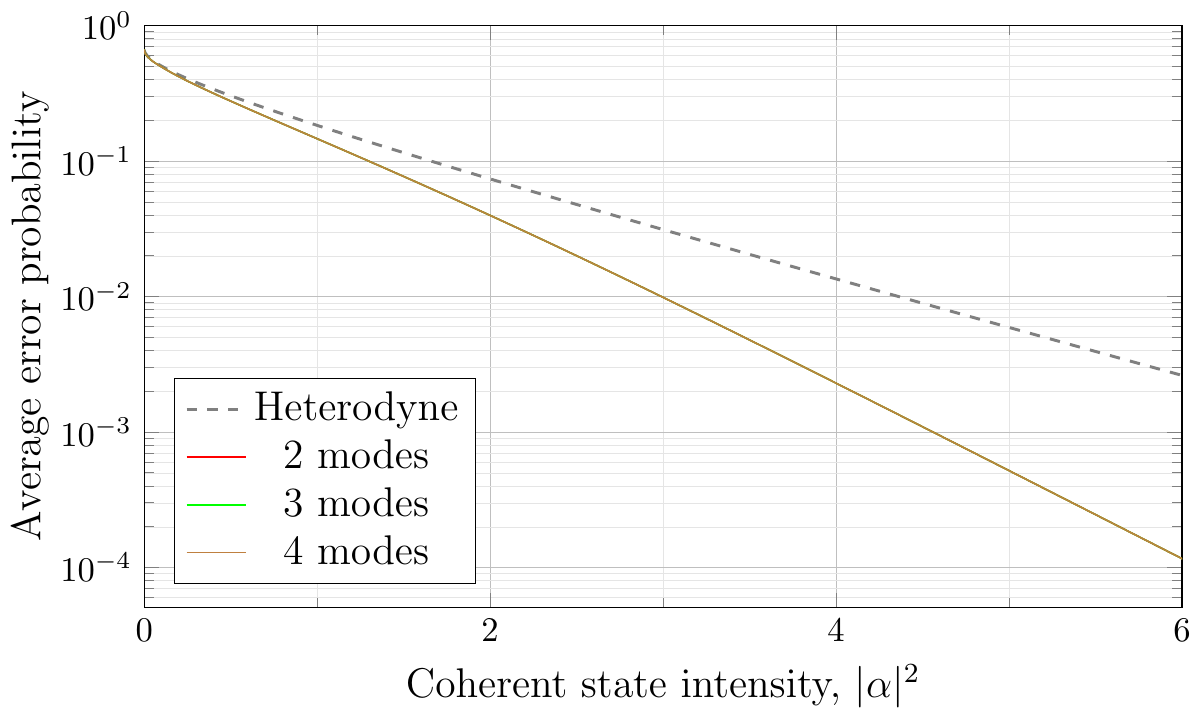}\label{fig:2psk}} \hspace{1pt}
\subfloat[5-PSK]{\includegraphics[width=0.326\linewidth]{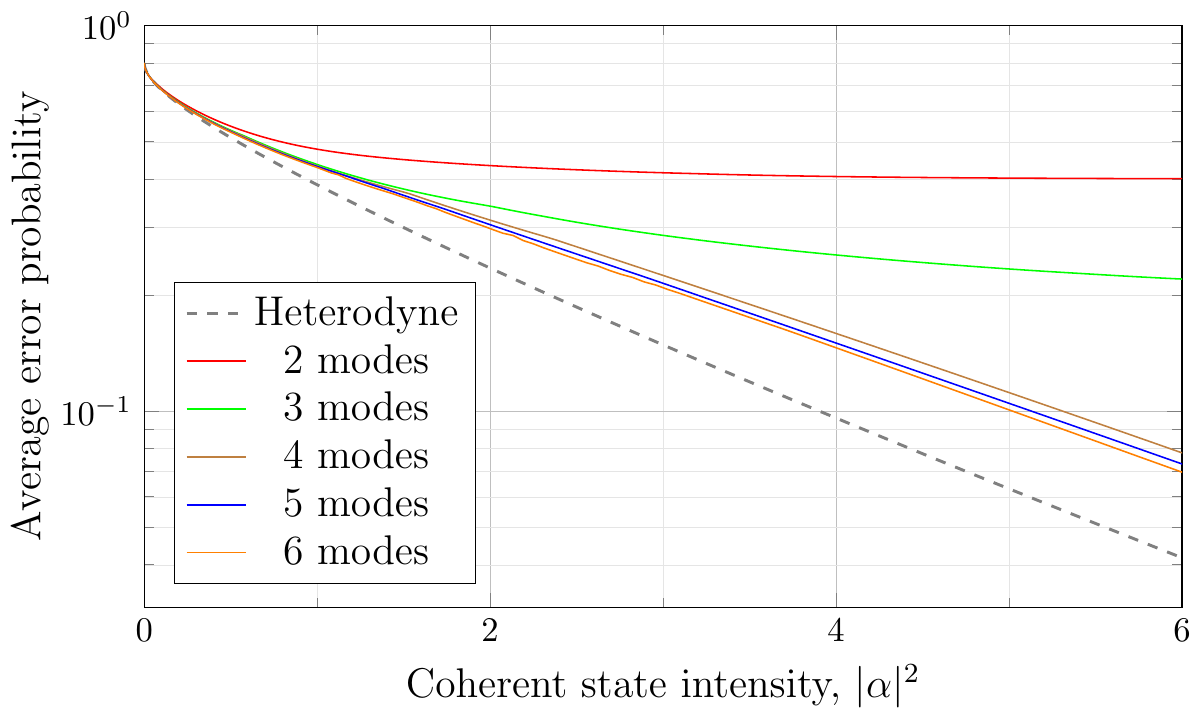} \label{fig:5psk}}\hspace{1pt}
\subfloat[6-PSK]{\includegraphics[width=0.326\linewidth]{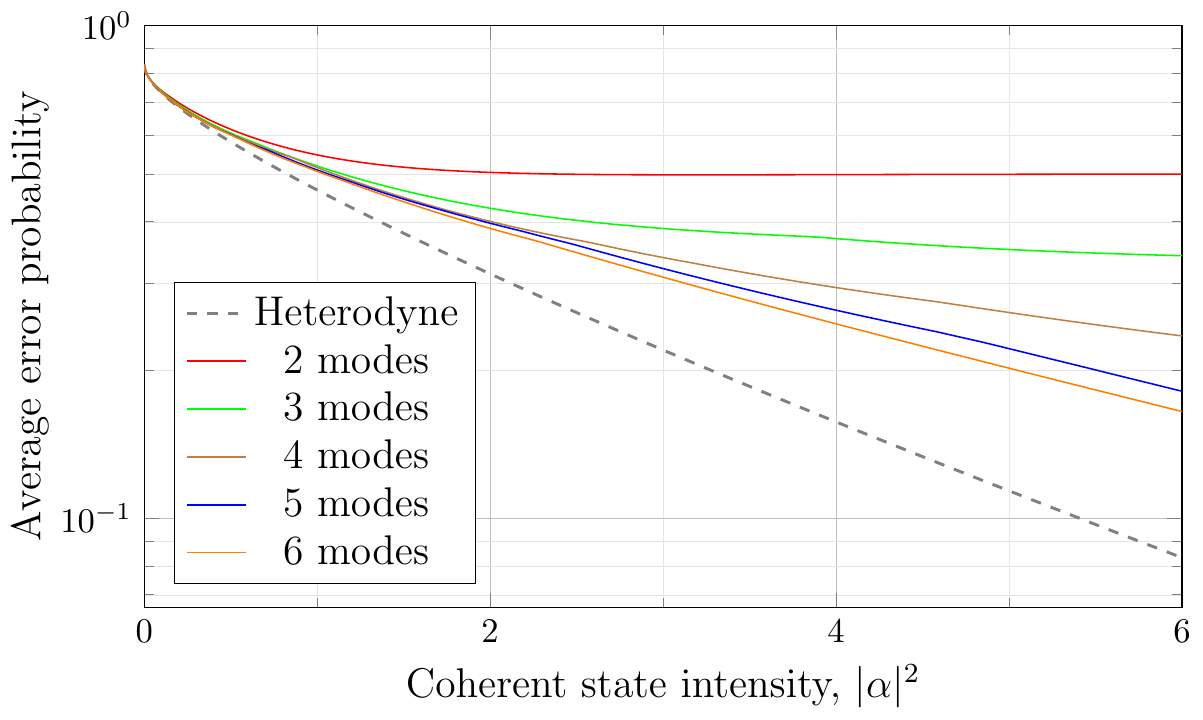} \label{fig:6psk}}
\caption{Average error rate for $M$-PSK signal discrimination with $M=\{3,5,6\}$ and varying number of ancillary modes. We apply our non-adaptive linear optical optimisation scheme and extend the discrimination beyond the regime of weak signals (where $\abs{\alpha}^2 > 1$).}
\label{fig:MPSK_Nmodes}
\end{figure*}%
%

\end{widetext}

\end{document}